\documentclass{article}
\usepackage{spconf,amsmath,graphicx,epsfig,epstopdf,subfigure}

\title{Human and Machine Speaker Recognition Based on Short Trivial Events}
\name{Miao Zhang$^{1}$, Xiaofei Kang$^{1}$, Yanqing Wang$^{1}$, Lantian Li$^{1}$, Zhiyuan Tang$^{1}$, Haisheng Dai$^{3}$, Dong Wang$^{1,2*}$\thanks{
This work was supported in part by the National Natural Science
Foundation of China under Projects 61371136 and 61633013,
and in part by the National Basic Research Program (973 Program) of China under
Grant 2013CB329302. Miao Zhang and Xiaofei Kang are joint first authors.
Corresponding author is Dong Wang (wangdong99@mails.tsinghua.edu.cn).} }
\address{1. Center for Speech and Language Technologies, RIIT, Tsinghua University \\
         2. Tsinghua National Laboratory for Information Science and Technology \\
         3. JD AI Research}
\begin{document}
%
\maketitle
\begin{abstract}

  Human speech often has events that we will call ¡°trivial events,¡± e.g., cough, laugh and sniff.
  Compared to regular speech, these trivial events are usually short and variable, thus
  generally regarded as not speaker discriminative and so are largely ignored by present
  speaker recognition research. However, these trivial events are highly valuable in
  some particular circumstances such as forensic examination, as they are less subjected
  to intentional change, so can be used to discover the genuine speaker
  from disguised speech.

  In this paper, we collect a trivial event speech database that involves $75$ speakers and
  $6$ types of events, and report preliminary
  speaker recognition results on this database, by both human listeners and machines. Particularly, the deep feature learning
  technique recently proposed by our group is utilized to analyze and recognize the trivial events, leading to
  acceptable equal error rates (EERs) ranging from $5\%$ to $15\%$ despite the extremely short durations (0.2-0.5 seconds) of these events. Comparing different types of events, `hmm' seems more speaker discriminative.

\end{abstract}
\section{Introduction}

Biometric authentication is highly important for the security of both reality and cyberspace.
Among various biometrics,
such as iris, palmprint, fingerprint and face, voiceprint has received much
attention recently, partly due to its convenience and
non-intrusiveness. After decades of research, $ $speaker recognition (SRE) by voiceprint
has achieved remarkable improvement.~\cite{unar2014review,kodituwakku2015biometric,beigi2011fundamentals,hansen2015speaker}

Most of the present SRE research works on `regular speech',
i.e., speech intentionally produced by people and involving clear linguistic content.
For this type of speech, rich speaker information can be obtained from both
vocal fold vibration and vocal tract modulation, so the speaker identifiablility
is generally acceptable. Many algorithms have been proposed to perform SRE
with this kind of speech, including the statistical model approach
that has gained the most popularity~\cite{Reynolds00,Kenny07,dehak2011front} and
the neural model approach that emerged recently and has attracted much
interest.~\cite{ehsan14,heigold2016end,li2017deep}

Despite the significant progress achieved on regular speech, research on the non-linguistic
part of speech signals is still very limited. For example, we may cough and laugh
when talking to others, and may `tsk-tsk'(people make with tongue when disapprove of something) or
`hmm'(people make to express doubt or uncertainty)
when listening to others. These events are produced by different personal habits and contain little linguistic information.
However, they do convey information about speakers. For example,
we can recognize a person by even a laugh if we have been familiar with him/her.
As these non-linguistic and non-regular events occur ubiquitously in our conversations,
we call them `trivial events'. Typical trivial events include
cough, laugh, `ahem'(short cough made by sb who is trying to get attention), etc.

A key value of SRE on trivial events is that these events are resistant
to potential disguise. In forensic examination, for example, the suspects may
intentionally change their voices to counteract the voiceprint testing, which
will largely fool the human listeners and cause failures in the existing SRE system. However, trivial events
are much harder to be counterfeited by the speaker, which makes it possible to
use these events to discover the true speaker from disguised speech. We will
show how disguised speech deceives humans and state-of-the-art SRE techniques in Section~\ref{sec:exp}.

An interesting question is: which type of trivial event conveys more speaker
information? Moreover, who is more apt to identify speakers from these events, human or machine?  In previous work, we have studied three
trivial events: cough, laugh and `wei' (Hello in Chinese), and found that
with a convolutional \& time-delay deep neural network (CT-DNN), an unexpected
high recognition accuracy can be obtained:
the equal error rate (EER) reaches as low as 11\% with
a cough of 0.3 seconds.~\cite{zhang2017speaker}
This good performance is largely attributed to the deep speaker feature learning
technique that we proposed recently.~\cite{li2017deep}

In this paper, we extend the previous work~\cite{zhang2017speaker} in several aspects:
(1) we extend the study to $6$ types of trivial events, i.e., cough, laugh, `hmm', `tsk-tsk', `ahem' and sniff;
(2) we collect a trivial event speech database and release it for public usage;
(3) we compare performance of human listeners and machines.

The organization of this paper is as follows:
the deep feature learning approach is briefly described in Section~\ref{sec:dfl},
and then the trivial event speech database CSLT-TRIVIAL-I is presented
in Section~\ref{sec:data}. The performance of human and machine tests
is reported in Section~\ref{sec:exp}, and some conclusions and discussions
are presented in Section~\ref{sec:con}.

\section{Related work}

Speaker recognition on trivial events is still limited. The most relevant work we
noticed is from Hansen et al.~\cite{hansen2017analysis,nandwana2014analysis} They analyzed the acoustic
properties of scream speech and studied the SRE performance on this
type of speech using a recognition system based on the Gaussian mixture
models-universal background model. Significant performance reduction was reported
compared with the performance on regular speech.

Some studies don't focus on trivial speech events we defined, but are still related to
our work. For example, Fan et al.~\cite{fan2011speaker} investigated the impact
of whisper speech on SRE, and Hanil{\c{c}}i et al.~\cite{hanilcci2013speaker} investigated
the impact of loud speech.

\section{Deep feature learning}
\label{sec:dfl}

Most of existing speaker recognition techniques are based on statistical models,
e.g., the Gaussian mixture model-universal
background model (GMM-UBM) framework~\cite{Reynolds00} and the subsequent subspace models,
such as the joint factor analysis approach~\cite{Kenny07}
and the i-vector model.~\cite{dehak2011front,lei2014novel} Additional gains
have been obtained by discriminative models and various normalization techniques (e.g., the SVM model~\cite{Campbell06}
and PLDA~\cite{Ioffe06}). A shared property of these statistical methods
is that they use raw acoustic features, e.g., the popular Mel frequency cepstral
coefficients (MFCC) feature, and rely on long speech segments to discover
the distributional patterns of individual speakers. Since most of trivial events are
short, these statistical models are not very suitable to represent them.

The neural model approach has gained much attention recently. Compared to the
statistical model approach, the neural approach focuses on learning
frame-level speaker features, hence more suitable for dealing with short
speech segments, e.g., trivial events. This approach was first proposed by Ehsan et al.~\cite{ehsan14},
where a regular deep neural network (DNN) was trained to discriminate
the speakers in the training data, conditioned on the input speech frames.
The frame-level features are then extracted from the last hidden layer, and an
utterance-based representation, called `d-vector', is derived by averaging the
frame-level features. Recently, we proposed a new convolutional \& time-delay
DNN (CT-DNN) structure, by which the quality of the learned speaker features
is significantly improved.~\cite{li2017deep} Particularly, we found that the new
features can achieve remarkable performance with short speech segments. This
property has been employed to recognize two trivial events (cough and laugh)
in our previous study, and good performance has been obtained.~\cite{zhang2017speaker}
More details about the CT-DNN model can be found in~\cite{li2017deep}, including the architecture and
optimization methods. The training recipe is also available online\footnote{http://project.cslt.org}.

In this paper, the deep feature learning approach will be mainly used to
recognize and analyze more trivial events, and the performance will be
compared with that obtained by human listeners.

%


\section{Database design}
\label{sec:data}

An appropriate speech corpus is the first concern before any analysis be conducted on trivial speech events.
Unfortunately, few trivial event databases are publicly available at present.
The only exception is the UT-NonSpeech corpus that was collected for
scream detection and recognition~\cite{hansen2017analysis,nandwana2014analysis},
but this corpus contains only screams, coughs and whistles. As we are more
interested in ubiquitous events that are not easy to be changed by
speakers intentionally, a more complicated database is required.
Therefore, we decided to construct our own database and release it
for public usage. This database is denoted by CSLT-TRIVIAL-I.

To collect the data, we designed a mobile application and
distributed it to people who agreed to participate.
The application asked the participants to utter $6$
types of trivial events in a random order, and each
event occurred $10$ times randomly. The random order
ensures a reasonable variance of the recordings for each event.
The sampling rate of the recordings was set to $16$ kHz and
the precision of the samples was $16$ bits.

We received recordings from $300$ participants.
The age of the participants ranges from $20$ to $60$,
and most of them are between $15$ and $30$.
These recordings were manually checked, and those recordings with clear
channel effect (noise, back-ground babbling and echo) were deleted.
Finally, the speech segments were purged and only a single event was
retained (e.g., one cough or one laugh) in each segment. After this
manual check, recordings from $75$ persons were remained, with $5$ to $10$ segments for each event per person.
Table~\ref{tab:TRIVIAL75} presents the data profile of
the purged database.

\begin{table}[htp]
    \begin{center}
        \caption{Data profile of CSLT-TRIVIAL-I}
        \label{tab:TRIVIAL75}
        \resizebox{\linewidth}{!}{
          \begin{tabular}{|l|c|c|c|c|}
           \hline
                          &   Spks    & Total Utts  &   Utts/Spk   &   Avg. duration (s)   \\
           \hline
               Cough      &  75	  & 732	&9.76	&   0.36  \\
           \hline
               Laugh     &   75	&709	&9.45 &0.39\\
           \hline
               `Hmm'      &  75 	&708	&9.44 &0.49\\
           \hline
               `Tsk-tsk'     &  75 	&1039	&13.85 &0.17\\
           \hline
               `Ahem'     &   75	&691	&9.21 &0.45\\
           \hline
               Sniff      &   75	&691	&9.21	   &0.37\\
           \hline
          \end{tabular}
       }
    \end{center}
\end{table}

Besides the trivial event database, we also collected
a disguise database. The goal of this database is to test how
human listeners and the existing SRE techniques will be
affected by speakers' intentional disguise.
This will provide a better
understanding about the value of our study on trivial events.

The same application used for collecting CSLT-TRIVIAL-I was used
to collect the recordings for the disguise database.
Before the recording,
the participants were instructed to try their best to counterfeit
their voices when recording the disguise speech.
During the recording, the application asked the participants to
pronounce $6$ sentences, each involving $5$ to $10$ words. Each sentence was spoken twice, one time
in the normal style and one time with intentional disguise.
In manual check, segments with much channel effect were removed.
After the manual check, recordings from $75$ speakers were
remained. This database is denoted by CSLT-DISGUISE-I.
Table~\ref{tab:DISGUISE75} presents the data profile in details.

CSLT-TRIVIAL-I and CSLT-DISGUISE-I have been released
online\footnote{http://data.cslt.org}. Users can download them freely
and use them under the Apache License Version $2.0$.

\begin{table}[htp]
    \begin{center}
        \caption{Data profile of CSLT-DISGUISE-I}
        \label{tab:DISGUISE75}
        \resizebox{\linewidth}{!}{
          \begin{tabular}{|l|c|c|c|c|}
           \hline
                          &   Spks    & Total Utts  &   Utts/Spk   &   Avg. duration (s)   \\
           \hline
               Normal     &   75   	  & 410	  & 5.47	 &  2.28    \\
           \hline
               Disguised  &   75      & 410	  & 5.47	 &  2.49 \\
           \hline
          \end{tabular}
       }
    \end{center}
\end{table}

\section{Experiments}
\label{sec:exp}

This section reports our experiments. We first present some details of two SRE systems
we built for the investigation, one based on the i-vector model and the other
based on the deep speaker feature learning
(denoted as d-vector system). Furthermore, performance with the two SRE systems on
CSLT-TRIVIAL-I is reported and compared with the performance of human listeners.
Finally, a disguise detection experiment conducted on CSLT-DISGUISE-I is reported,
which demonstrates how speech disguise fools both humans and the existing SRE systems.

\subsection{SRE systems}

For the purpose of comparison, we build two SRE systems, an i-vector system and a d-vector system.
For the i-vector system, the input feature involves $19$-dimensional MFCCs plus the log energy,
augmented by its first and second order derivatives.
The UBM is composed of $2,048$ Gaussian components, and the dimensionality of
the i-vector space is $400$. Three scoring methods are used: cosine distance,
cosine distance after LDA projection, and PLDA.
The dimensionality of the LDA projection space is $150$. When PLDA is used for scoring, the i-vectors are
length-normalized. The system is trained using the Kaldi SRE08 recipe.~\cite{povey2011kaldi}

For the d-vector system, the input feature involves $40$-dimensional Filter banks(Fbanks). A symmetric $4$-frame window is used to
connect the neighboring frames, resulting in $9$ frames in total.
The number of output units is $5,000$, corresponding to the number of speakers in
the training data.
The frame-level speaker features are extracted from the last hidden layer,
and the d-vector of each utterance is derived by averaging all its frame-level speaker features.
The scoring methods used for the i-vector system are also used for the
d-vector system during the test, including cosine distance, LDA and PLDA.

The Speech-ocean Datatang database is used as the training set, which was recorded by telephone and the sampling rate
is $16$ kHz. The database consists of $5,000$ speakers, with $80,3654$ Chinese utterances. This training set is used to train the UBM, the T matrix,
and the LDA/PLDA models of the i-vector system, as well as the CT-DNN model of the d-vector system.

\subsection{SRE on trivial events}

In the first experiment, we evaluate the SRE performance on trivial events, by both human listeners
and the two SRE systems.
The CSLT-TRIVIAL-I database is used to conduct the test. It consists of $75$ speakers and $6$ types of trivial
events, each type per speaker involving about $10$ segments. The original data of the recording is in $16$ kHz, which matches the Speech-ocean Datatang database.

During the human test, the listener is presented $36$ YES/NO questions,
$6$ questions per event type. For each question, the listener is asked to listen to two speech segments that
are randomly sampled from the same event type, with a probability of 50\% to be from the same speaker.
Listeners are allowed to perform the test multiple times. We collected $33$
test sessions, amounting to $1,188$ trials in total. The performance is
evaluated in terms of detection error rate (DER), which is the proportion of the incorrect answers
within the whole trials, including both false alarms and false rejections.
The results are shown in Table~\ref{tab:trivial-h}. It can be seen that humans can
tell the speaker from a very short trivial event, particularly with the nasal sound `hmm'.
For cough, laugh and `ahem', humans can obtain some speaker information, but the performance is lower.
For `tsk-tsk' and sniff, the performance is very bad, and the answers given by the listeners are almost random. This is
expected to some extent, as these two types of events sound rather weak, and producing them does
not use much of vocal fold and vocal tract.

    \begin{table}[htb]
    \begin{center}
      \caption{DER of human test on trivial events.}
      \label{tab:trivial-h}
      \scalebox{1}{ 
          \begin{tabular}{|c|c|c|c|c|c|}
            \hline
                     \multicolumn{6}{|c|}{DER\%}\\
            \hline
                Cough   &  Laugh  &  `Hmm'    & `Tsk-tsk'   & `Ahem'    &  Sniff       \\
            \hline
                20.20   &  20.71  &  19.70   & 42.42   & 26.26    & 35.86       \\
           \hline
          \end{tabular}
      }
      \end{center}

   \end{table}

For the machine test, there are about $260,000$ trials for each event type.
The EER results with the i-vector system and the d-vector system are
reported in Table~\ref{tab:trivial}.
It can be observed that the d-vector system outperforms the i-vector system by a
large margin, confirming that the deep speaker feature learning approach is more suitable than
the statistical model approach when recognizing short speech segments. Comparing
different events, it can be found that `hmm' conveys the most speaker information, and cough,
laugh, `ahem' are less informative. `Tsk-tsk' and Sniff are the least discriminative. All these observations
are consistent with the results of the human test. Moreover, we found that for d-vector systems, the discriminative
normalization approaches, LDA and PLDA, did not provide clear advantage on 'hmm' and sniff. A possible reason is that
there is little intra-speaker variances involved in these two types of events, so the statistical based discrimination is not helpful.

Comparing humans and machines, we can find that the best machine system, i.e., the d-vector system,
is highly competitive. Although DER and EER values are not directly comparable, the results still
show roughly that on almost all the types of trivial events, the d-vector system makes fewer mistakes
than humans. Particularly, on the events that humans perform the worst, i.e., `tsk-tsk' and sniff, machines work
much better. Although the listeners we invited are not professional speech scientists, and the results may be
affected by the audio devices that human listeners used, these results still provide strong
evidence that machines can potentially do better than human beings in listening to trivial events.

    \begin{table}[htb]
    \begin{center}
      \caption{EER results on CSLT-TRIVIAL-I with the i-vector and d-vector systems.}
      \label{tab:trivial}
      \resizebox{1\linewidth}{!}{          
          \begin{tabular}{|l|l|c|c|c|c|c|c|}
            \hline
            \multicolumn{2}{|c|}{}                 &\multicolumn{6}{c|}{EER\%}\\
            \hline
               Systems & Metric & Cough   &  Laugh  &  `Hmm'    & `Tsk-tsk'   & `Ahem'    &  Sniff       \\
            \hline
               i-vector& Cosine & 23.42   &  27.69  & 15.71   & 29.70   & 18.12   &  37.78     \\
                       & LDA    & 26.14   &  27.99  & 15.54   & 31.79   & 20.83   &  37.74   \\
                       & PLDA   & 27.82   &  25.79  & 14.28   & 33.57   & 21.85   &  34.76     \\
           \hline
               d-vector& Cosine & 8.89         &  12.43         & \textbf{5.88} & 16.75          & 10.44          &  \textbf{11.91}    \\
                       & LDA    & \textbf{8.33}&  \textbf{11.20}& 6.76          & \textbf{15.95} & \textbf{9.71}  &  12.44   \\
                       & PLDA   & 10.26        &  15.48         & 7.28          & 17.85          & 13.16          &  12.93    \\
           \hline
          \end{tabular}
      }
      \end{center}

   \end{table}



\subsection{Disguise detection}

In the second experiment, we examine how humans and machines can discriminate disguised speech.
For the human test, the listener is presented $6$ trials, each containing
two samples from the same speaker, but one of the sample can be a disguised version.
The listener is asked to tell if the two samples are from the same speaker. To avoid any bias,
the listeners are informed that some speech samples are disguised. Some trials may also involve
imposter speech (not the same speaker), but these trials are only used to inject noise into the
test, not counted in the final result.
We collected $198$ trails in total, and the DER result is $47.47\%$. This indicates that
human listeners largely fail in discriminating disguised speech.

The EER results of the two SRE systems are reported in Table~\ref{tab:disguise}. It can
be found that machines can do better than humans in discriminating disguised speech, but the
error rates are still very high. Again, the d-vector system performs better than the i-vector system.

    \begin{table}[htb]
    \begin{center}
      \caption{EER results on CSLT-DISGUISE-I with the i-vector and d-vector systems.}
    \vspace{2mm}
      \label{tab:disguise}
      \resizebox{0.5\linewidth}{!}{          
          \begin{tabular}{|l|c|c|}
            \hline
            \multicolumn{1}{|c|}{}                 &\multicolumn{2}{c|}{EER\%}\\
            \hline
               Metric & i-vector&  d-vector         \\
            \hline
               Cosine & 28.70    &  25.74        \\
               LDA    & 34.57    &  \textbf{24.17}       \\
               PLDA   & 28.70    &  28.17        \\
           \hline
          \end{tabular}
      }
      \end{center}
   \end{table}

\begin{figure}
    \centering
    \includegraphics[width=0.75\linewidth]{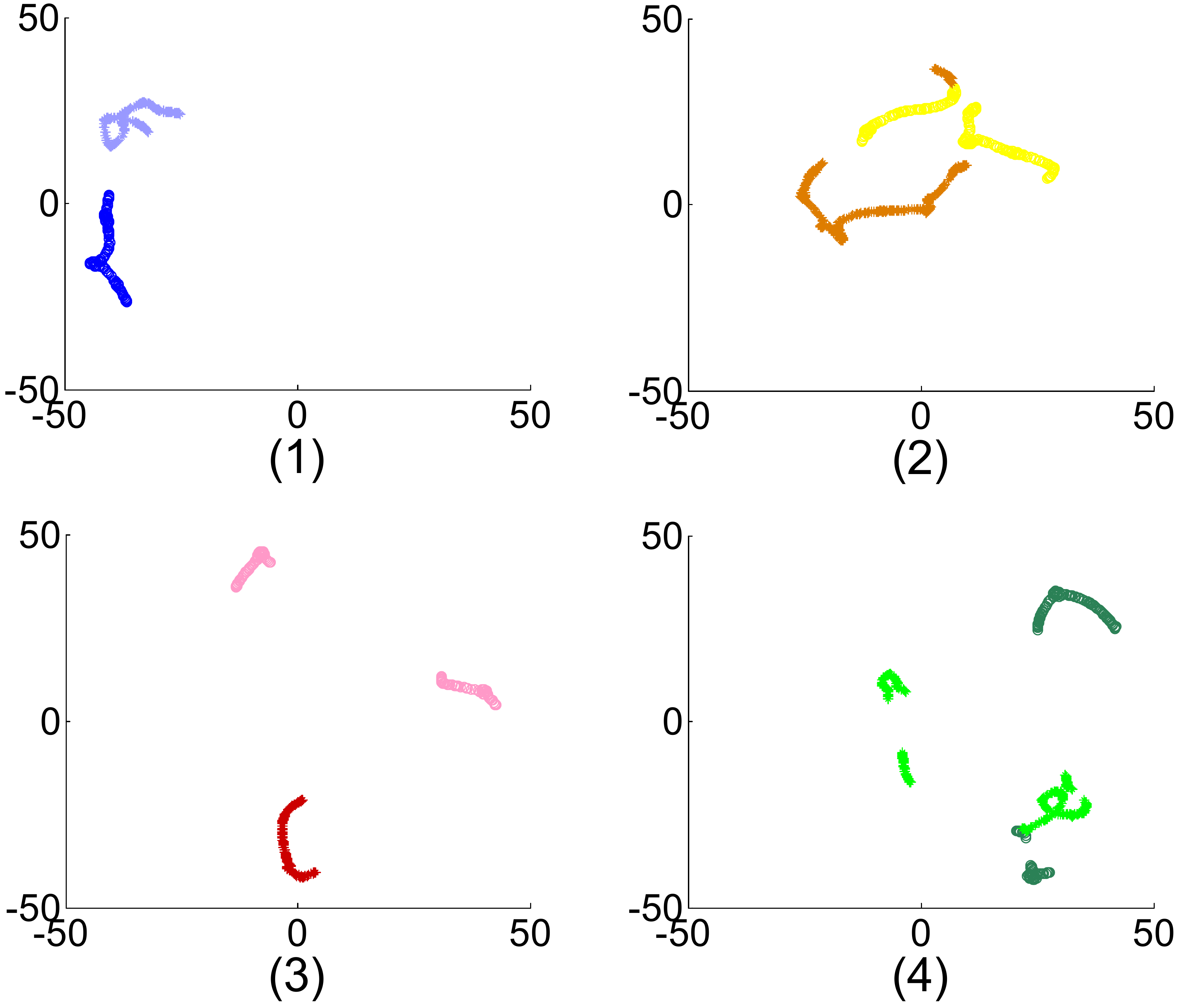}
    \caption{The deep speaker features of the normal speech and disguised speech from the same speaker and the same
    sentence plotted by t-SNE. Each picture represents a single person, and normal and disguised speech are represented by darker and
    lighter curves, respectively.}
    \label{fig:disguise}
    \end{figure}

To observe the impact of speech disguise more intuitively, we plot the deep speaker features produced
by the d-vector system in 2-dimensional space using t-SNE.~\cite{saaten2008} The results are shown in Fig.~\ref{fig:disguise}.
We can see that the discrepancy between the normal and disguised speech is highly speaker-dependent: some speakers are not
good voice counterfeiters, but some speakers can do it very well.

\section{Conclusions}
\label{sec:con}

In this paper, we studied and compared the performance of human listeners and machines
on the speaker recognition task with trivial speech events. Our experiments on $6$ types of
trivial events demonstrated that both humans and machines can discriminate speakers to some extent with
trivial events, particularly those events involving clear vocal tract activities,
e.g., `hmm'. Additionally, the deep speaker feature learning approach works much better than the
conventional statistical model approach on this task, and in most cases outperforms
human listeners. We also tested the performance of humans and machines on disguised speech,
and found that speech disguise does place a serious challenge for both of them.

\newpage
\bibliographystyle{IEEEbib}
\bibliography{refs}

\end{document}